\documentclass[aps,prb,reprint,superscriptaddress]{revtex4-1}

\usepackage{amssymb,amsfonts,amsmath}
\usepackage{bm}
\usepackage[applemac]{inputenc}
\usepackage{paralist} 
\usepackage{pdfsync} 
\usepackage{color}
\usepackage{graphicx}
\def\Mw{$M_{\textrm{w}}$}

\begin{document}


\title{Multi-phase microstructures drive exciton dissociation in neat semicrystalline polymeric semiconductors}


\author{Francis~Paquin}
\affiliation{D\'epartement de physique \& Regroupement qu\'ebécois sur les mat\'eriaux de pointe, Universit\'e de Montr\'eal, C.P.\ 6128, Succursale centre-ville, Montr\'eal (Qu\'ebec) H3C~3J7, Canada}

\author{Jonathan~Rivnay}
\author{Alberto~Salleo}
\affiliation{Materials Science and Engineering,  Stanford University, 476 Lomita Mall, 239 McCullough Building, Stanford, California 94305, United~States}

\author{Natalie~Stingelin}
\affiliation{Department of Materials and Centre for Plastic Electronics, Imperial College London, South Kensington Campus, London SW7~2AZ, United~Kingdom}

\author{Carlos~Silva-Acu\~na}
\email[Corresponsing author. E-mail: ]{carlos.silva@umontreal.ca}
\affiliation{D\'epartement de physique \& Regroupement qu\'ebécois sur les mat\'eriaux de pointe, Universit\'e de Montr\'eal, C.P.\ 6128, Succursale centre-ville, Montr\'eal (Qu\'ebec) H3C~3J7, Canada}
\affiliation{Visiting Professor, Experimental Solid State Physics, Department of Physics, Imperial College London, South Kensington Campus, London SW7~2AZ, United~Kingdom}


\date{\today}

\begin{abstract}
The optoelectronic properties of macromolecular semiconductors depend fundamentally on their solid-state microstructure and phase morphology. Hence, it is of central importance to manipulate --- from the outset --- the molecular arrangement and packing of this special class of polymers from the nano- to the micrometer scale when they are integrated in thin film devices such as photovoltaic cells, transistors or light-emitting diodes, for example.  One effective strategy for this purpose is to vary their molecular weight. The reason for this is that materials of different weight-average molecular weight (\Mw) lead to different microstructures. Polymers of low \Mw\:form unconnected, extended-chain crystals because of their non-entangled nature. As a result, a polycrystalline, one-phase morphology is obtained. In contrast, high-\Mw\:materials, in which average chain lengths are longer than the length between entanglements, form two-phase morphologies comprised of crystalline moieties embedded in largely un-ordered (amorphous) regions. Here, we discuss how changes in these structural features affect exciton dissociation processes. We utilise neat regioregular poly(3-hexylthiophene) (P3HT) of varying \Mw\:as a model system and apply time-resolved photoluminescence (PL) spectroscopy to probe the electronic landscape in a range of P3HT thin-film architectures. We find that at 10\,K, PL originating from recombination of long-lived charge pairs decays over microsecond timescales. Tellingly, both the amplitude and decay-rate distribution depend strongly on \Mw. In films with dominant one-phase, chain-extended microstructures, the delayed PL is suppressed as a result of a diminished yield of photoinduced charges. Its decay is significantly slower than in two-phase microstructures. We therefore conclude that excitons in disordered regions between crystalline and amorphous phases dissociate extrinsically with yield and spatial distribution that depend intimately upon microstructure, in agreement with previous work [Paquin~et~al.\ \textit{Phys.\ Rev.\ Lett.}, 2011, \textbf{106}, 197401]. We note, however, that independent of \Mw, the delayed-PL lineshape due to charge recombination is representative of that in low-\Mw\:microstructures. We thus hypothesize that charge recombination at these low temperatures --- and likely also charge generation --- occur in torsionally disordered chains forming more strongly coupled photophysical aggregates than those in the steady-state ensemble, producing a delayed PL lineshape reminiscent of that in paraffinic morphologies at steady state.
\end{abstract}


\maketitle

\section{Introduction}

Understanding how the solid-state microstructure of neat polymeric semiconductors influences their photophysical properties is of fundamental importance for many plastic optoelectronic devices. For example, understanding of such inter-relationship may assist in gaining insights in organic solar cells, which generally rely on so-called bulk heterojunctions, in which electron acceptors (typically fullerene derivatives) are blended with a $\pi$-conjugated polymer, which acts as the electron donor.~\cite{Clarke:2010gb} The resulting solid-state microstructures can be highly complex, typically involving multiple phases that consist of crystalline and amorphous domains that are rich in each component, and potentially intermixed phases, such as amorphous solutions comprised of molecularly mixed donor and acceptor molecules, or, more rarely, ordered co-crystals composed of the two components.~\cite{Treat:2010im,Collins:2010hf,Jamieson:2012hr} 
This structural complexity has rendered elucidation of relevant details of the disordered energy landscape that drives electronic processes such as charge separation challenging. 

\begin{figure}[th]
 \centering
 \includegraphics[width=\columnwidth]{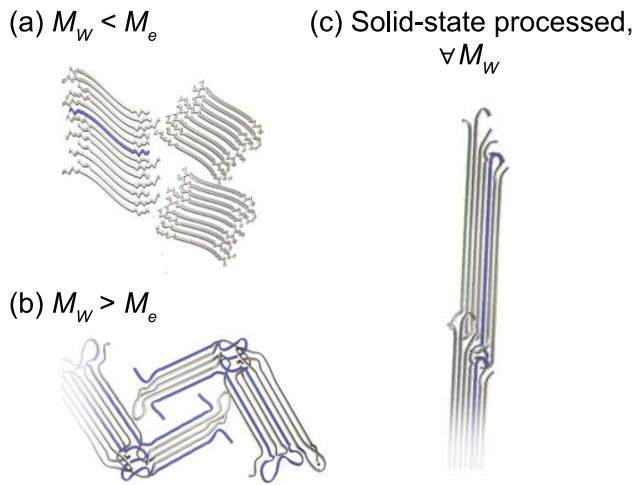}%
 \caption{Schematic of the different solid-state microstructures produced by the P3HT samples used in this work. (a) P3HT films of $M_{\textrm{w}} \leqslant 40$\,kg/mol lead to isolated, unconnected extended-chain crystals, forming polycrystalline one-phase morphologies. In this regime, the chain length is shorter than that correponding to the molecular weight, $M_e$, over which chain entanglement occurs. Any particular oligomer chain (e.g.\ that highlighted in blue) is part of only one crystal on average. (b) P3HT of $M_{\textrm{w}} \geqslant 50$\,kg/mol, where chain lengths exceed $M_e$, leads to a two-phase microstructure, in which crystalline lamella are embedded in amorphous regions. This structure results from chain entanglements in the liquid state, which hinder molecular ordering during solidification.\cite{Wunderlich:1976} Individual polymer chains may serve as tie molecules connecting different crystals through amorphous regions (highlighted in blue). (c) By pressing the polymer powder at temperatures below the melting point $T_m$, the lamellar crystal thickness is enlarged, thereby inducing predominantly chain-extended morphologies that depend weakly on \Mw.~\cite{Baklar:2012} \label{fig:processing}}
 \end{figure}

In this article, we address the relationship between the polymer microstructure and electronic dynamics by focusing on exciton dissociation and recombination processes in neat, regioregular P3HT. In a previous letter,\cite{Paquin:2011} we reported that tightly bound geminate-polaron pairs are formed on sub-picosecond timescales following ultrafast photoexcitation, which subsequently recombine by tunnelling over distributed timescales, producing delayed PL. We speculated that these are generated in intermediate regions between crystalline and amorphous phases, and that the recombination dynamics depend upon the local energy landscape in these regions. In order to develop this hypothesis, we present here time-resolved PL measurements in neat P3HT films of \Mw\:ranging over 12--348\,kg/mol. For \Mw\;$\lesssim40$\,kg/mol, polycrystalline one-phase morphologies dominate in solution-processed films (Fig.~\ref{fig:processing}(a)).\cite{Brinkmann:2007,Brinkmann:2011,Koch:2013} For similar processing, material of \Mw\;$\gtrsim50$\,kg/mol produces complex two-phase architectures, in which crystalline lamellae are embedded in largely amorphous regions (Fig.~\ref{fig:processing}(b)).\cite{Wunderlich:1976} Independent of \Mw, predominantly chain-extended microstructures are induced by pressing the polymer in the solid state (Fig.~\ref{fig:processing}(c)).\cite{Jog:1999, Baklar:2012} As a consequence, by choice of molecular weight and processing route, solid-state microstructures ranging from one-phase polycrystalline through semicrystalline (two-phase) P3HT architectures can be prepared. 
We utilise this approach to obtain information on how certain structural features affect the electronic landscape in this interesting class of materials, and find that indeed exciton dissociation is favoured by the disordered landscape at gradual interfaces between crystalline and amorphous regions, with a yield of photoinduced charges that depend on the solid-state microstructure. Furthermore, as the disorder along the $\pi$--$\pi$ stacking direction increases, the mean electron-hole separation increases, leading to a slower distribution of charge recombination times. Our conclusions, using neat P3HT, help to shed light on how structural and energetic disorder plays a role in driving excitonic processes in more complex systems, such as the important bulk heterojunction structures. More generally, our work provides a general fundamental link between organic-semiconductor photophysics and classical polymer science. For instance, insights gained may be adapted to establish a more refined picture of the structural evolution of bulk commodity polymers, where spectroscopic methods used for conjugated macromolecules cannot be applied.

\section{Experimental Methods}

 \subsection{Polymer processing}
P3HT films of weight-average molecular weight in the range of 12, 50, 200, 270, and 350\,kg/mol were investigated. All materials had a similar regioregularity and a polydispersity of, respectively, 1.7, 1.8, 2.6, 2.1, and 3.7. 
Samples were processed in two different ways. They were wire-bar-coated from p-xylene solution (1.4\% by wt) on glass substrates. The solution and substrate temperature was 70\,$^{\circ}$C. Throughout this paper, we refer to samples processed this way as ``solution-processed'' films. Alternatively, they were solid-state pressed from powder as described elsewhere;\cite{Baklar:2012} these samples are described here as ``solid-state-processed'' films.

\subsection{Photoluminescence spectroscopy}
Steady-state PL measurements were carried out with a continuous-wave laser (Ultralasers Inc., 200\,mW maximum, 532\,nm) and chopped at 100\,Hz with a mechanical chopper (Terahertz technologies). The sample was positioned inside a sample-in-exchange-gas, closed-cycle cryostat (Cryoindustries of America) and kept at 10\,K. The PL was coupled to a 300-mm spectrometer (SP2300i, Princeton Instruments)  and detected with a Si/PbS photoreceiver (S/PBS-025/020-TE2-H, Electro-Optical Systems Inc). The signal from the photoreceiver and the synchronization signal from the chopper controller were sent to a lock-in amplifier (SR830, Stanford Research Systems). PL spectra were collected by scanning the spectrometer, and were corrected for the instrument spectral response of the experimental apparatus. 
Time-resolved PL measurements were carried out with a 40-fs, 532-nm (2.33-eV) pulse train derived from an optical parametric amplifier (Light Conversion TOPAS), pumped by a Ti:sapphire laser system (KMLabs Dragon, 1-kHz repetition rate). Maximum pump fluences were 5\,$\mu$J\,cm$^{-2}$. Spectra were measured with an intensified CCD camera (Princeton-Instruments PIMAX 1024HB) coupled to a 300-mm spectrometer (SP2300i, Princeton Instruments), with a 5-ns electronic gate synchronized to the pulse train. Time-resolved PL spectra were obtained by varying the gate delay and temporal width electronically with respect to the laser-pulse arrival time. All time-resolved PL spectra were corrected for the instrument response of our spectrometer.

\subsection{Wide-angle X-ray scattering}
X-ray scattering experiments were performed at the Stanford Synchrotron Radiation Lightsource (SSRL) on beamline 7-2 (high-resolution grazing incidence), with an  incident energy of 8\,keV. The diffracted beam was collimated with 1\,mrad Soller slits for high-resolution in-plane scattering. For grazing incidence experiments, the incidence angle was slightly larger than the critical angle, ensuring that we sampled the full film depth. All synchrotron X-ray scattering measurements were performed under a He atmosphere to reduce air scattering and beam damage to the sample. Full Warren Averbach peak shape analysis and single peak estimates of grazing incidence diffraction peaks were performed as detailed elsewhere.\cite{Rivnay:2011,Noriega:2013aa}

\section{Results and analysis}

\begin{figure}[th]
		\includegraphics [width=\columnwidth]{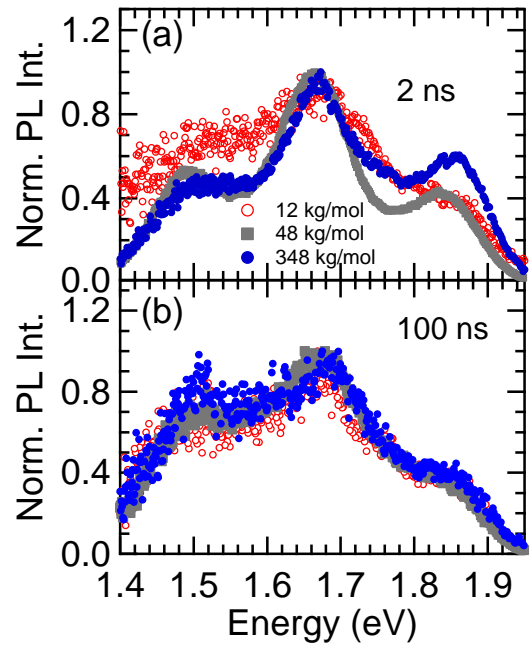}
		\caption{Time-gated photoluminescence spectra measured at 10\,K. Prompt (a) and delayed (b) spectra for solution-processed films of different weight-average molecular weight, indicated in the legend of panel a, are displayed. The time delay after excitation is indicated in each panel. In both panels, the spectra were normalised to have equal intensity at 1.70\,eV, the energy of the 0--1 vibronic peak.
	\label{fig:TRPLspec}}
\end{figure}
\begin{figure}[th]
		\includegraphics [width=\columnwidth]{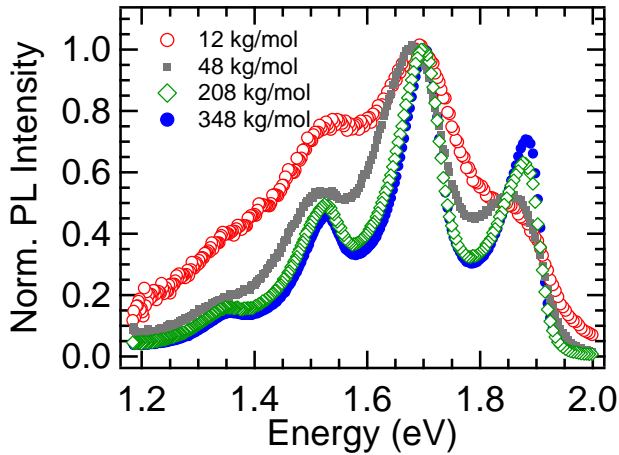}
		\caption{Steady-state photoluminescence spectra measured at 10\,K for solution-processed films of different weight-average molecular weight, indicated in the legend. The spectra were normalised to have equal intensity at 1.70\,eV, the energy of the 0--1 vibronic peak.
	\label{fig:SSPLspec}}
\end{figure}

Fig.~\ref{fig:TRPLspec} shows time-resolved PL spectra for solution-processed films at time delays indicated in each panel. 
With the shortest temporal gate of our detector (5\,ns, Fig.~\ref{fig:TRPLspec}(a)), these films display spectral lineshapes that depend on \Mw. Over this short gate window, we measure the integrated PL intensity over the exciton intrinsic lifetime in neat P3HT.\cite{Paquin:2011} Because this component is the dominant emission, the early-time gated PL spectra resemble the steady-state PL spectra, which are shown in Fig.~\ref{fig:SSPLspec}. 

As \Mw\:increases, two clear trends in both the prompt and steady-state PL spectral lineshapes are evident:
\begin{inparaenum}[(\itshape i\upshape)]
\item the relative amplitude of the 0--0 peak at 1.88\,eV, when spectra are normalized to the 0--1 feature at 1.70\,eV, increases at high \Mw;
\item  the effective Huang-Rhys parameter, here defined as $\lambda^2_{\textrm{eff}} = 2 I^{0-2}/I^{0-1}$,  decreases from $\lambda^2_{\textrm{eff}} = 1.5 \pm 0.2$ at \Mw\;$ = 12$\,kg/mol to $\lambda^2_{\textrm{eff}} = 0.9 \pm 0.1$ at \Mw\;$ = 348$\,kg/mol.  
\end{inparaenum}
In a separate publication, we have reported steady-state PL spectra of solution-processed P3HT thoughout this range of \Mw, similar to the spectra shown in Fig.~\ref{fig:SSPLspec}, and modelled the spectral lineshape evolution with \Mw\:within the framework of a photophysical aggregate model taking into account both intra- and intermolecular electronic dispersion.\cite{Paquin:2013} The 0--0/0--1 ratio depends entirely  on the intrachain excitonic (resonance-Coulomb) coupling,\cite{Clark:2007,Spano:2009,Spano:2014vn} and $\lambda^2_{\textrm{eff}}$ reflects the degree of torsional disorder along the chain backbone.\cite{Paquin:2013}  

Using such a photophysical aggregate model, we have reported that the interchain excitonic coupling, characterised by the 0--0/0--1 absorbance ratio, decreases abruptly for materials with $M_w \geq 50$\,kg/mol.\cite{Reid:2012,Paquin:2013} This coincides with a transition from a one-phase microstructure (Fig.~\ref{fig:processing}(a)) to a two-phase system (Fig.~\ref{fig:processing}(b)).\cite{Brinkmann:2007,Brinkmann:2011,Koch:2013} Importantly, this abrupt transition in excitonic coupling around \Mw\:$\sim 50$\,kg/mol influences the exciton coherence length derived from steady-state PL spectra:\cite{Paquin:2013} in the low-\Mw\:regime, the exciton coherence length is larger along the $\pi$-stack direction than in the high-\Mw\:regime. Conversely, as \Mw\:increases, polymer chains constituting the lamellar lattice are on average more planar as they are sterically less influenced by chain ends, and more susceptible to neighbouring chains in the $\pi$ stack.\cite{Koch:2013} Therefore, the decrease of $\lambda^2_{\textrm{eff}}$ and the increase of the 0--0/0--1 PL ratio with increasing \Mw\:is consistent with longer, more torsionally ordered, planar chains with decreasing interchain excitonic coupling. The time-resolved PL spectral lineshapes in Fig.~\ref{fig:TRPLspec}(a) reproduce the steady-state PL specra reported in Fig.~\ref{fig:SSPLspec},\cite{Paquin:2013} and we interpret this trend accordingly.

On longer time windows, we observe delayed PL (Fig.~\ref{fig:TRPLspec}(b)), which can in principle arise from triplet bimolecular annihilation dynamics, but that we assigned in ref.~\citenum{Paquin:2011} to recombination of geminate charge pairs over distributed timescales by considering its fluence dependence as well as due to the lack of triplet photoinduced absorption signatures in quasi-steady-state photoinduced absorbtion measurements. Furthermore, linear fluence dependence over more than one order of magnitude to lower fluences~\cite{Paquin:2011} rules out extrinsic charge generation by two-step excitation events.~\cite{Silva:2001nx,Silva:2002nr}  In all of the samples, the relative 0--0 PL intensity is lower than that on short time windows, and it depends weakly on \Mw. In fact, in all samples, the relative 0--0 intensity resembles that in the steady-state spectrum of films of the lowest \Mw\:(Fig.~\ref{fig:SSPLspec}). Furthermore, the Huang Rhys parameter is similar in the delayed PL spectra of all samples, at $\lambda^2_{\textrm{eff}} \sim 1.5$, which is again reminiscent of that in the lowest \Mw\:samples at early time (Fig.~\ref{fig:TRPLspec}(a)) and at steady state (Fig.~\ref{fig:SSPLspec}).\cite{Paquin:2013} These observations suggest that slow charge recombination events populate a sub-ensemble that is \emph{always} characterized by a structural features similar to the ones found in the average bulk of films with the lowest \Mw, characterized by higher interchain excitonic coupling and higher intrachain torsional disorder than the broader ensemble of emitters over the fast gate window, resulting in longer interchain spatial coherence and more limited intrachain one. This points to a \Mw-independence of the microstructure hosting charge recombination, which we have argued is the same environment as the charge separation environment at 10\,K.\cite{Paquin:2011} We therefore conclude that charge separation occurs in regions with highest configurational disorder, which resemble low-\Mw\:average microstructures, in all samples. 


\begin{figure}[th]
		\includegraphics [width=\columnwidth]{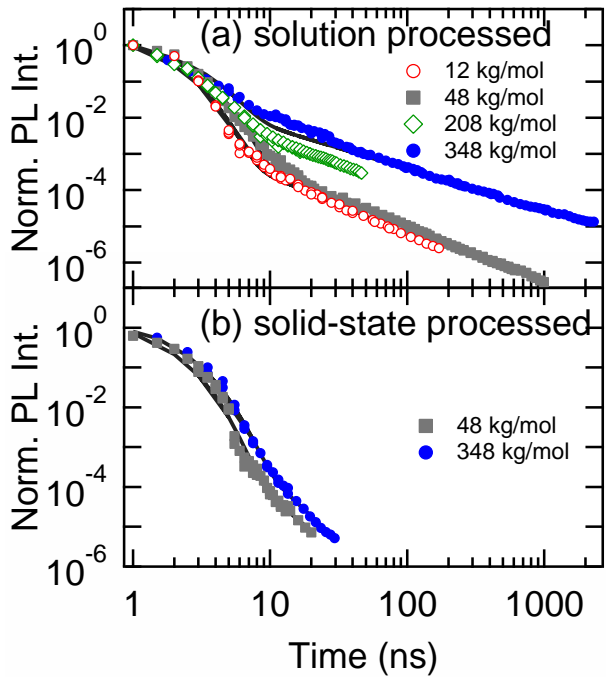}
		\caption{Time-resolved photoluminescence intensity measured at 10\,K. The weight-average molecular weight is indicated in the figure for both solution (a) and solid-state processing conditions (b). The intensity was determined by integrating time-resolved spectra such as those displayed in Fig.~\ref{fig:TRPLspec}. The data were normalised to have equal intensity at a 1-ns delay. 
	\label{fig:PLdecay}}
\end{figure}

In order to quantify differences in geminate-pair generation and recombination kinetics, we consider the evolution of the spectrally integrated PL intensity. Fig.~\ref{fig:PLdecay}(a) shows time-resolved PL intensity of solution-processed films of different \Mw, measured at 10\,K. 
We have previously assigned the long power-law component to distributed recombination of geminate-polaron pairs.\cite{Paquin:2011} Here, we observe a marked dependence of the power-law decay kinetics on \Mw; the long-lived emission amplitude increases significantly ($>36$\% of total PL intensity) for samples with \Mw\;$=348$\,kg/mol, relative to that in ref.~\cite{Paquin:2011} (\Mw\;$=48$\,kg/mol; $>12$\% delayed PL intensity). The increased delayed PL intensity with increasing \Mw\:is accompanied by a slow-down of the power-law decay. We note that power-law component of PL decay for lower molecular weight samples is very similar to the weak power law decay characteristics observed in isolated P3HT nanofibers of similar molecular weight.~\cite{Labastide:2012fk}

Tellingly, while the power-law decay kinetics are dramatically different in the diverse solid-state microstructures arising from solution processing, they are weakly dependent on \Mw\:in the solid-state processed films (Fig.~\ref{fig:PLdecay}(b)). These show a significant decrease in the amplitude of the power law for all \Mw\:studied, as well as an increase of the decay rate with respect to solution-processed films. 
We underline that in the solid-state processed samples, the surface area of photophysical aggregate and non-aggregate phases is reduced in all samples regardless of molecular weight (see Fig.~\ref{fig:processing}(c)), leading to a substantial reduction in the geminate pair yield, as the volume of conformationally disordered regions is reduced substantially.

\begin{figure}[t]
		\includegraphics[width=\columnwidth] {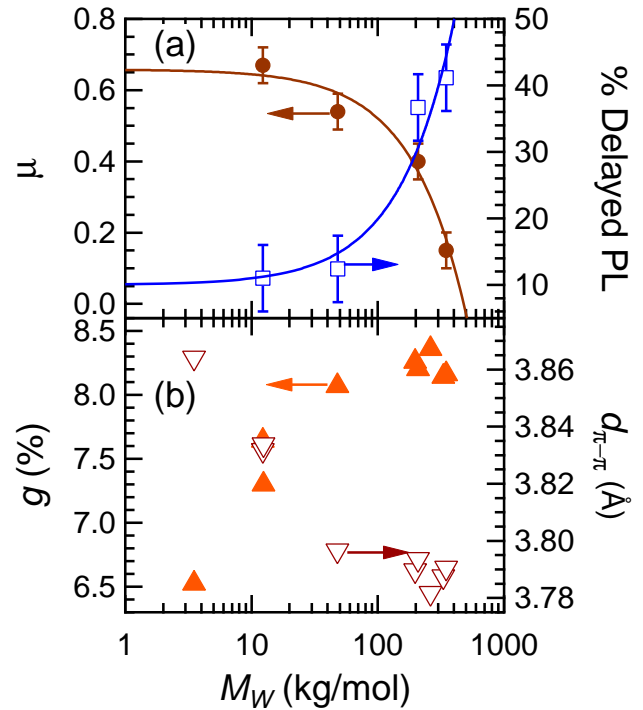}
		\caption{(a) Power-law decay parameter $\mu$ (left axis) and delayed PL relative intensity (right axis) versus \Mw\:in solution-processed films. The parameters are obtained by fitting the long-time time-resolved PL data such as that displayed in Fig.~\ref{fig:PLdecay} to a power law in the form $I \propto t^{1+\mu}$, with the delayed PL intensity corresponding to the power-law amplitude. Within the model invoked here, $\mu$ represents the ratio of the characteristic geminate-pair radii and the distance dependence of the charge recombination rate. A linear fit to $\mu$(\Mw) has intercept $0.66 \pm 0.03$ and slope $(-1.37 \pm 0.16) \times 10^{-3}$\,mol/kg. 
		(b)~Paracrystalline disorder $g$ and $\pi$--$\pi$ stacking distance $d_{\pi-\pi}$ versus \Mw, derived from wide-angle X-ray scattering measurements. Disorder in $d_{\pi-\pi}$ is quantified by $g$.
	\label{fig:mu}}
\end{figure}

We now consider the evolution of the slope of the power-law decay with \Mw\, in greater detail. To explain the time-resolved PL intensity decay, we have invoked in a previous publication a model which accounts for rapid branching between bright excitons and dark geminate polaron pairs following photoexcitation, followed by a distribution of times for recombination.\cite{Paquin:2011} The initial exponential component thus arises from relaxation of the emissive state during the excited-state lifetime, while the power-law component results from regeneration of excitons via recombination of geminate polarons with a distribution of rates. The absence of any marked temperature dependence in the power-law decay, as demonstrated in ref.~\citenum{Paquin:2011}, suggests that recombination of geminate pairs occurs via tunnelling of charges on distinct polymer chains, such that the recombination rate constant is exponentially distance-dependent, $k(r) \propto e^{-\beta r}$. Furthermore, following Tachiya and Seki,\cite{Tachiya:2009} we assumed an exponential distribution of geminate-pair separation for the purposes of modelling, $f(r) = \epsilon e^{-\epsilon r}$. Within this model, the waiting time distribution for recombination is determined by these spatial functions, $R(t) = \int_{0}^{\infty} f(r) k(r) e^{-k(r)} dr$. The long-time PL decay is governed by geminate-pair recombination events, and is functionally of the form $I (t\gg \tau) \propto t^{1+\mu}$, where $\tau$ is the excited-state lifetime and $\mu \equiv \epsilon /\beta$. Thus, if the characteristic electron-hole separation $\epsilon$ in the distribution $f(r)$ increases, or if $\beta$ decreases, $\mu$ increases and the power-law decays more slowly. We show the dependence of $\mu$ on \Mw\:in the solution-processed films in Fig.~\ref{fig:mu}(a), which shows a marked apparently linear decrease with increasing \Mw.

In order to explore the trend in $\mu$ by considering the evolution of lamellar packing with \Mw, 
we have calculated the paracrystalline disorder parameter, $g$, in the $\pi$-stacking direction, which is the standard deviation in the $\pi$-stacking distance as a percentage of its mean.\cite{Hindeleht:1988,Rivnay:2011} Therefore, $g_{\pi}$ is directly related to structural disorder in the inter-chain distance along the stack. Through peak shape analysis of the principal and higher order $\pi$-stacking diffraction peaks, deconvolution of the effects of cumulative lattice disorder and finite crystallite size is possible using a model based on that of Warren and Averbach. 
This allows us to rank structural disorder quantitatively, from a perfectly ordered crystalline lattice ($g = 0$\%) to an amorphous one ($g \geq \sim10$\%). It has been shown that for semicrystalline polymers, especially those of high molecular weight, the influence of paracrystalline disorder ($g$) dominates the peak shape, thus allowing for single peak-width estimation of $g$.\cite{Rivnay:2011} Fig.~\ref{fig:mu}(b) shows an increase in $g$ with \Mw\:from $\sim6.5$\% to $>8$\%. 
Clearly, as \Mw\:increases, the $\pi$-stacking order decreases, and consequently the average electron-hole separation increases, leading to a broader distribution of electron-hole recombination rates and a slower power-law PL decay. Another factor that contributes to this development is the increase in interface area between crystalline and amorphous phases, where the disordered energy landscape is more complex than in one-phase systems.

It is now useful to attempt to quantify the characteristic electron-hole separation in different samples based on Fig.~\ref{fig:mu}. We do not have a reliable measurement of $\beta$ in this class of materials, but assuming that $\beta \sim 1$\,\AA$^{-1}$, which is typical for charge tunnelling in molecular systems,\cite{Barbara:1996} then characteristic geminate-pair radii vary from $\epsilon^{-1}=1.5$\,\AA~for films made from 12\,kg/mol materials, to 6.7\,\AA~for those fabricated from 348\,kg/mol material. For $\pi$--$\pi$ stacking distances in these samples, $\sim 3.8$\,\AA, this implies that the average electron-hole separation is less than one nearest-neighbour distance in the lowest \Mw\:samples, and more than one nearest-neighbour distance in the highest \Mw\:samples. The paracrystalline disorder thus correlates with the average radius and yield of geminate-polaron pairs, underlining the influence of structural disorder and exciton dissociation processes.

\section{Discussion}

The key results of the previous section are that
\begin{inparaenum}[(\itshape i\upshape)]
\item the \emph{amplitude} of delayed PL contribution to the integrated PL intensity increases systematically with increasing \Mw~in solution-processed films (Fig.~\ref{fig:PLdecay}(a)), but drops dramatically in solid-state-processed films independent of \Mw~(Fig.~\ref{fig:PLdecay}(b));\label{BigRes1}
\item although both the prompt PL spectral lineshape (Fig.~\ref{fig:TRPLspec}(a)) and the delayed PL decay kinetics (Fig.~\ref{fig:PLdecay}(a)) depend unambiguously on \Mw\:in solution-processed films, there is a striking independence of the delayed PL spectral lineshape on $M_w$ (Fig.~\ref{fig:TRPLspec}(b))\label{BigRes2}. 
\end{inparaenum}
The platform to discuss these observations is our preliminary report in which we concluded that geminate polaron pairs are generated on fast timescales with respect to the exciton lifetime, without significant exciton diffusion, and that these recombine in a distribution of times by quantum tunnelling in the same region in which they were generated.\cite{Paquin:2011} Observation~(\textit{\ref{BigRes1}}) establishes that as solution-processed solid-state microstructures evolve from a polycrystalline morphology (Fig.~\ref{fig:processing}(a)) to a two-phase one featuring a higher fraction of chain-entangled configurations (Fig.~\ref{fig:processing}(b)), the yield of photoinduced geminate polaron pairs increases, and the radial distribution (the distance between the centre of mass of the electron and hole) increases as well. When the microstructure is forced to be predominantly chain-extended by solid-state processing (Fig.~\ref{fig:processing}(c)), both the yield and radius distribution of geminate polarons decreases dramatically for all values of \Mw. Furthermore, observation~(\textit{\ref{BigRes2}}) establishes that while the contribution of amorphous phases is important for the charge generation yield, states that are re-populated by geminate-pair recombination are characteristic of photophysical aggregates in the long- and/or short-ranged ordered phases, but these aggregates are often defined by a high degree of torsional disorder and high average interchain excitonic coupling, evident from the spectral lineshape of delayed PL.\cite{Paquin:2013} Therefore, we conclude that the polymer conformations that constitute regions between moderately ordered domains and amorphous phases are key in the geminate polaron pair photogeneration mechanisms in neat P3HT. This microstructure dependence on the yield and radius distribution of charge photogeneration that we demonstrate in this communication is likely to be general to semicrystalline semiconductor polymer films. In fact, Labastide et~al.\ showed that in P3HT nanoparticles formed by processing from aqueous solution show a marked dependence of delayed PL dynamics with particle size,\cite{Labastide:20112} with $\mu$ \emph{increasing} with increasing particle diameter, suggesting important microstructure evolution with particle spatial characteristics.

A fundamentally important question that emerges from this interpretation is: what is the mechanism of this fast branching between self-trapped excitons and geminate polaron pairs in these complex microstroscures? A starting consideration is that as this is a neat $\pi$-conjugated polymer system and as such is characterized by highly bound Frenkel excitons at equilibrium, there should be no significant driving force for exciton dissociation in the absence of molecular electron acceptors such as fullerene derivatives or chemical defects. In fact, at steady state, the interchain exciton coherence length across the entire \Mw\:range studied here varies by only $\sim 20$\%.\cite{Paquin:2013} The vibrationally dressed exciton density, that is, the electronic excitation that is largely localised on a single chain with the vibrational lattice distortion dressing it, spans 2--4 chains at equilibrium, depending on \Mw. Nevertheless, polaron signatures have also been reported in P3HT neat films,~\cite{Hendry:2004p1534,Ai:2006p6883,Sheng:2007,Cunningham:2008p6673,Piris:2009} nanofibres,~\cite{Labastide:2012fk} and nanoparticles~\cite{Labastide:20112} by diverse groups. A possible explanation is that nascent excitons experience substantially longer spatial coherence across different chains on ultrafast timescales.\cite{Banerji:2013} This phenomenon has been invoked to account for ultrafast charge separation in polymer heterostructures.\cite{Banerji:2011,Hermann:2011,Kaake:2012} On ultrashort ($\ll 100$\,fs) timescales, bath-induced quantum coherence between Frenkel-exciton and delocalized charge-transfer states is possible,\cite{Bittner:2013} and its dynamics would be correlated to the dynamic motion of the molecular framework.\cite{Rozzi:2013fk} Transient coherent photoexcitations would branch into the self-trapped excitons inferred from steady-state PL spectra,\cite{Paquin:2013} or into geminate polaron pairs, on ultrafast timescales corresponding to decoherence dynamics. The excitation spectrum of delayed PL demonstrates that exciting into states that do not constitute photophysical aggregates in the ground state ($>2.4$\,eV) enhances the delayed PL yield,\cite{Paquin:2011} and the role of delocalized charge-transfer states coherently coupled to exciton states could be the origin of that phenomenon. Here, the excitation photon energy employed in our measurements is 2.33\,eV, which is close to the isosbestic point between photophysical aggregate and non-aggregate absorption.\cite{Clark:2009} Previous ultrafast PL studies on P3HT films have pointed out that even upon excitation with $\sim 3$-eV photon energy, well into the non-aggregate absorption range,~\cite{Clark:2007} the transient PL spectrum is characteristic of photophysical aggregates within the experimental instrument response, typically $\sim 100$\,fs.\cite{Banerji:2011} This indicates that even upon excitation well into energy ranges in which chromophores do not `feel' significant intermolecular coupling in the ground state, photoexcitations rapidly relax into photophysical aggregates, on timescales much faster than vibrational relaxation of the chains (whose signatures are evident from transient PL spectra on $>10$-ps timescales.\cite{Parkinson:2010}) Hot photoexcitations are argued by some to play a crucial role in driving photocarrier generation at polymer donor-acceptor heterojunctions used in photovoltaic diodes.\cite{Jailaubekov:2013,Grancini:2013} 
Similarly, we argued in ref.~\citenum{Paquin:2011}, and we maintain here, that excitons with excess energy in disordered, complex multi-phase P3HT films yield charge pairs on ultrafast timescales in these neat samples. Note however, that as demonstrated and discussed in ref.~\citenum{Paquin:2011}, we  generate tightly bound geminate polaron pairs in neat P3HT, not photocarriers as in bulk heterostructures optimised for polymer-based solar cells. 
Here we only speculate that the nature of the transient coherent excitation following the first few femtoseconds after light absorption, and the local chain conformations hosting them, might be what drives branching into predominantly charge-transfer-like excitations. \emph{Direct} spectroscopic probes of spatial coherence of these highly transient states remains an important challenge for the ultrafast spectroscopy community concerned with the photophysics of plastic semiconductors, but may be probed by multidimensional spectroscopies.\cite{Collini:2009}

Steady-state PL spectral lineshapes in P3HT films are very well described by a hybrid HJ-aggregate model, introduced by Yamagata and Spano,~\cite{Yamagata:2012jk} over the range of \Mw\:used in this work.~\cite{Paquin:2013}  In that model, the hybrid HJ coupling is composed of Mott-Wannier (intrachain) and Frenkel (intrerchain) excitonic  coupling, and charge-transfer (CT) interactions. The sign and magnitude of the CT contribution is expected to be highly sensitive to interchain order.  We presume these interactions to give rise to interchain excitons in crystalline domains which can either decay radiatively, as discussed extensively by Labastide~et~al.,\cite{Labastide:2012fk} or separate into polaron pairs. In isolated crystalline fibers, this branching ratio is essentially unity in favor of excitons,\cite{Labastide:2012fk} similar to the solid-state processed samples discussed in this work. We consider that this branching depends on the total area of gradual interface between crystalline and phases dominated by chain entanglements, which depends strongly on \Mw. 

Perhaps a more banal but plausible explanation of the microstructure dependence of delayed PL phenomena reported here involves chemical defects, likely arising from photo-oxidation of P3HT. Majewski et~al.\ have reported that degradation of field-effect transistors based on P3HT show marked kinetics depending on processing conditions, and attributed the differences to the extent to which water and oxygen diffuse to crystalline/amorphous interfaces due to differences in microstructure.\cite{Majewski:2006db} As the volume fraction of amorphous regions of the film increases with \Mw, it is possible that the density of dopants at interfaces also increases, producing a higher amplitude of delayed PL in our samples. We cannot rule out this possibility, although it is unclear why the decay kinetics would depend on \Mw\:in this scenario. However, even if dopant-induced quenching accounts to a substantial extent for the delayed PL dynamics reported here, the same fundamental semiconductor polymer science issues stand: there is a clear relationship between solid-state microstructure and exciton quenching dynamics, and two-phase microstructures enhance this process due to the complex energy landscape intrinsic to these disordered architectures. This complexity can then be generalized to binary systems involving electron acceptor moieties.

Reid et~al.\ reported time-resolved microwave conductivity measurements on the same \Mw\:series and processing routes (Fig.~\ref{fig:processing}) as reported here,\cite{Reid:2012} allowing direct comparison with this work. Those measurements were carried out at room temperature and probe photocarriers, in contrast to the tightly-bound charge pairs probed by our delayed PL measurements at 10\,K.\cite{Paquin:2011} Their data show that the yield of photocarriers increases with increasing molecular weight in the paraffinic regime (Fig.~\ref{fig:processing}(a)) and saturates at the transition into the entangled, semicrystalline regime (Fig.~\ref{fig:processing}(b)). The main conclusion was that the evolution of semicrystalline microstructure with well-defined interfaces between amorphous and crystalline polymer domains is necessary for spatial separation of the electron and hole, which controls the yield of free charges. We consider that there is a general conclusion to be drawn from both types of optical probes involving the importance of solid-state microstructure on electronic dynamics in semicrystalline organic semiconductors. The development of chain configurations defining two-phase semicrystalline microstructures define a complex disordered energetic landscape that infuences the branching of transient photoexcitations into charge separated states. 

Generally, in classical polymer science, understanding of the complex architectures formed by high-\Mw\:macromolecules is essential because many of the properties of bulk commodity ``plastics'' are dictated by two-phase morphologies of long-chain polymers. The development of mechanical properties with \Mw\:is fundamentally no different in P3HT than in common plastics, and electronic properties such as charge transport display a corresponding development.\cite{Koch:2013} Here we have demonstrated that excitonic processes in neat P3HT are also profoundly dependent on microstructure via \Mw\:and processing routes, and we consider that this is a general property of plastic semiconductors.

\section{Conclusions}

By means of time-resolved PL spectroscopy at 10\,K, we have explored geminate-polaron-pair yields and their recombination dynamics in films composed of neat P3HT of varying \Mw, and processed either from solution or in the solid state. Over the range of \Mw\:studied and the distinct processing used, the solid-state microstructure varies from being polycrystalline (one-phase) to semicrystalline (two-phase crystalline/amorphous). The yield of photogenerated geminate polaron pairs, their distribution of radii (the separation between electron and hole), and their distribution of recombination rates depends sensitively on the nature of the microstructure. These findings illustrate that the key consideration to determine electronic properties in semicrystalline polymeric semiconductors is to control the solid-state microstructure by appropriate choice of molecular weight and by processing routes.




\providecommand*{\mcitethebibliography}{\thebibliography}
\csname @ifundefined\endcsname{endmcitethebibliography}
{\let\endmcitethebibliography\endthebibliography}{}

\end{document}